\documentclass[twocolumn,showpacs,amsfonts,aps,prc,floatfix,superscriptaddress,nofootinbib]{revtex4}

\usepackage{amsmath}
\usepackage{bm}
\usepackage{graphicx}

\newcommand{\be}[1]{\begin{equation}\label{#1}}
\newcommand{\ee}{\end{equation}}

\usepackage{color}
\usepackage{soul}

\begin{document}
\setstcolor{blue}

%%%%%%%%%%%%%%%%%%%%%%%%Front Matter%%%%%%%%%%%%%%%%%%%%%%%%%%%%%%%%%%
%%%%%%%%%%%%%%%%%%%%%%%%%%%%%%%%%%%%%%%%%%%%%%%%%%%%%%%%%%%%%%%%%%%%%%

\title{Spectra and elliptic flow for identified hadrons in 2.76 A TeV Pb+Pb collisions}
\author{Huichao Song}
\email[Correspond to\ ]{Huichaosong@pku.edu.cn}
\affiliation{Department of Physics and State Key Laboratory of Nuclear Physics and Technology, Peking
University, Beijing 100871, China }
\affiliation{Collaborative Innovation Center of Quantum Matter, Beijing 100871, China}

\author{Steffen A. Bass}
\affiliation{Department of Physics, Duke University, Durham,
             North Carolina 27708, USA}

\author{Ulrich Heinz}
%\email[Email:\ ]{heinz@mps.ohio-state.edu}
\affiliation{Department of Physics, The Ohio State University,
%  191 West Woodruff Avenue,
  Columbus, Ohio 43210-1117, USA}

\begin{abstract}
Using  the {\tt VISHNU} hybrid model that couples (2+1)-dimensional viscous hydrodynamics to a microscopic hadronic transport model, we calculate the multiplicity, $p_T$ spectra and elliptic flow for pions, kaons, and protons in 2.76 A TeV Pb+Pb collisions, using MC-KLN initializations with smoothed initial conditions, obtained by averaging over a large number of events. The results from our calculations are compared to data from the ALICE collaboration, showing nice agreement over several centrality bins. Using the same inputs, we predict the $p_T$ spectra and elliptic flow for $\phi$ mesons and explore its flow development in the strong and weak coupling limits through hydrodynamic calculations with different decoupling temperatures. In addition we study the influence of baryon and anti-baryon annihilation processes on common observables and demonstrate that by including  annihilation processes below a switching temperature of 165 MeV, {\tt VISHNU} provides a good description of the multiplicity and $p_T$-spectra for pions, kaons and protons measured by PHENIX and ALICE at both RHIC and the LHC.
\end{abstract}

\date{\today}

\maketitle

%%%%%%%%%%%% begin text %%%%%%%%%%%%%%%%%%%
\section{Introduction}\label{sec1}
\vspace*{-3mm}

In relativistic heavy ion collisions at top Relativistic Heavy Ion Collider (RHIC) and  Large Hadron Collider (LHC) energies, more than 99\% of the hadrons are produced with transverse momenta below 2\,GeV, and the quark-gluon plasma (QGP) matter produced in these collisions behaves as an almost perfect liquid~\cite{Rev-Arsene:2004fa,Gyulassy:2004vg,Muller:2006ee,reviews}. Local pressure gradients convert the initial fireball deformations and inhomogeneities into fluid momentum anisotropies, which then translate into  flow harmonics that describe the asymmetry of particle productions in momentum space~\cite{reviews,Heinz:2013th,Gale:2013da}. The shear viscosity of the fluid controls the conversion efficiency, which leads to a suppression of elliptic flow and higher order flow coefficients as discovered by different groups~\cite{Romatschke:2007mq,Song:2007fn,Dusling:2007gi,Molnar:2008xj,Bozek:2009dw,Chaudhuri:2009hj,Xu:2007jv,Schenke:2010rr,Qiu:2011hf}.

The integrated elliptic flow of all charged hadrons $v_2^{ch}$ has been used  to extract the specific shear viscosity of the QGP $(\eta/s)_{QGP}$ since it is directly related to the momentum anisotropy of the fluid and monotonically decreases with $(\eta/s)_{QGP}$~\cite{Song:2010mg,Song:2011hk}. On the other hand, the differential elliptic flow $v_2(p_T)$ for identified hadrons heavily depends on the chemical composition and radial flow of the system during hadronic stage of the reaction evolution. As a result, $v_2(p_T)$ for identified hadrons is more sensitive to the details of the theoretical calculation, yet it can also be used to test the extracted QGP viscosity obtained from the integrated $v_2$ for all charged hadrons.

Using the {\tt VISHNU} hybrid model~\cite{Song:2010aq} that connects the hydrodynamic expansion of the viscous QGP fluid to the microscopic kinetic evolution of the hadronic matter, we previously extracted the QGP viscosity from the integrated elliptic flow for all charged hadrons in 200 A GeV Au+Au collisions and provided bounds on  $(\eta/s)_{QGP}$, $ 1{\,<\,}4\pi(\eta/s)_{QGP}{\,<\,}2.5$, where the uncertainties were dominated by the initial condition models~\cite{Song:2010mg}. Within that extracted QGP viscosity range, {\tt VISHNU} was able to provide an excellent description of all soft-hadron data at top RHIC energy~\cite{Song:2011hk}. After extrapolating the calculations to 2.76 A TeV Pb + Pb collisions, we demonstrated the same for the elliptic flow data for all charged hadrons measured by the ALICE collaboration  with approximately the same or slightly higher value of the specific QGP viscosity~\cite{Song:2011qa}. However, so far the  data for identified hadrons at the LHC have not been fully explored within the {\tt VISHNU} hybrid approach, except for a report on preliminary results for $v_2(p_T)$ for pions, kaons and protons in~\cite{Heinz:2011kt}.

This article investigates in detail spectra and elliptic flow of identified soft hadrons in 2.76\,$A$\,TeV Pb+Pb collisions. It is organized as follows: Sec.~\ref{sec2} briefly introduces the {\tt VISHNU} hybrid model and the set-up for our calculations. Sec.~\ref{sec3} studies the centrality dependence of the multiplicity, $p_T$ spectra and differential elliptic flow for pions, kaons, and protons and compares these to data taken by the ALICE collaboration. Using the same parameters, we then predict the $p_T$ spectra and elliptic flow for $\phi$ mesons in Sec.~\ref{sec4} and subsequently explore its flow development in the strong and weak coupling limits through hydrodynamic calculations with different decoupling temperatures. In Sec.~\ref{sec5} we investigate the influence of $B{-}\bar{B}$ annihilation processes on soft particle production and show that with proper inclusion of $B{-}\bar{B}$ annihilation processes and a switching temperature of $T_\mathrm{sw}{\,=\,}165 \ \mathrm{MeV}$, {\tt VISHNU} can nicely reproduce the multiplicity and spectra for pions, kaons and protons at both, RHIC and LHC energies. A short summary is presented in Sec.~\ref{sec6}.

%%%%%%%%%%%%%%%%%%%%%%%%%%%%%%%%%%%%%%%%%%%%%%%%%
\section{Setup of the Calculation}\label{sec2}
\vspace*{-3mm}

In this article, we utilize the {\tt VISHNU} hybrid model~\cite{Song:2010aq} to investigate identified soft hadron productions in 2.76 TeV Pb+Pb collisions. {\tt VISHNU} connects the 2+1-d relativistic viscous hydrodynamic model ({\tt VISH2+1})~\cite{Song:2007fn} for the QGP fluid expansion to the microscopic hadronic transport model ({\tt UrQMD})~\cite{Bass:1998ca} for the description of hadron re-scattering and the evolution of the hadron gas. Using a modified Cooper-Frye formula that accounts for viscous corrections,  a Monte-Carlo event generator ({\tt H2O}) converts the hydrodynamic output into an ensemble of hadrons for propagation in the microscopic transport model.\footnote{The particles emitted from the 2+1-d fluid are boost-invariant distributions. The {\tt H2O} event generator uses this boost invariance to extend the 2+1-d hydrodynamic output at $y{\,=\,}\eta_s{\,=\,}0$ to non-zero momentum and space rapidities, and then samples the particle momentum distributions within the rapidity range $|y|<4$. In spite of edge effects near the forward and backward ends of this rapidity window, the 3+1-d {\tt UrQMD} evolution retains longitudinal boost invariance around mid-rapidity in the range $|y|<1.5$~\cite{Song:2010aq}.}
The default switching temperature $T_{\mathrm{sw}}$ between the macroscopic and microscopic approaches in {\tt VISHNU} is set to 165\,MeV, which is close to the QCD phase transition temperature \cite{Aoki:2006br,Bazavov:2011nk}. For the hydrodynamic evolution above $T_{\mathrm{sw}}$, the default equation of state (EOS) utilized is {\tt s95p-PCE} which has been constructed by matching lattice QCD data at high temperature to a chemically frozen hadron resonance gas (with chemical decoupling temperature $T_\mathrm{chem}{\,=\,}165$\,MeV) at low temperature \cite{Huovinen:2009yb}. Following Ref.~\cite{Song:2011qa}, initial entropy density profiles are generated using the MC-KLN model~\cite{Drescher:2006ca,Hirano:2009ah}, by averaging over a large number of fluctuating entropy density distributions (individually recentered and aligned with the reaction plane). The choice of initial conditions directly affects the hydrodynamic flow and thus the extracted value of the QGP specific shear viscosity. However, it does not directly influence the relative difference among the distributions of the momentum anisotropy of the different hadron species related to the mass ordering of $v_2$. Since this article does not aim to extract QGP viscosity at the LHC with reliable uncertainty estimates, but extends the previous investigations for all charged hadrons to identified hadrons, we simply follow~\cite{Song:2011qa} and use the MC-KLN initialization.

For simplicity, we neglect net baryon density, heat flow, and bulk viscosity~\cite{Song:2009rh}. The QGP specific shear viscosity $(\eta/s)_\mathrm{QGP}$  is assumed to be a constant and the corresponding relaxation time is set to $\tau_{\pi}=3\eta/(sT)$~\cite{Song:2007fn}. Ref.~\cite{Song:2011qa} found that, in order to fit the ALICE integrated and differential elliptic flow $v_2\{4\}$ data for all charged hadrons, $(\eta/s)_\mathrm{QGP}$ ranges from 0.20 to 0.24.  In this article, $(\eta/s)_\mathrm{QGP}$ is reduced to 0.16 for a better description of new $v_2$ data for pions, kaons and protons measured by the scalar product method~\cite{Noferini:2012ps}. Non-flow and fluctuation effects contaminate the different flow measurements differently~\cite{Ollitrault:2009ie,Voloshin:2008dg}, leading to the slightly different $(\eta/s)_\mathrm{QGP}$ values used in this article vs. earlier work. This will be further explained in the next section. For $(\eta/s)_\mathrm{QGP}=0.16$, the initial time is set to $\tau_0 =0.9 \ \mathrm{fm/c}$, obtained from fitting the slope of the $p_T$ spectra for all charged hadrons below 2 GeV~\cite{Song:2011qa}.

%%%%%%%%%%%%%%%%%%%%%%%%%%%%%%%%%%%%%%%%%%%%%%%%%
\section{Multiplicity, spectra and elliptic flow for identified hadrons}\label{sec3}

%%%%%%%%%%%%%%%% Fig. 1 %%%%%%%%%%%%%%%%%%%%%%%%%%%%%%
\begin{figure}[t!]
\includegraphics[width=0.85\linewidth,clip=]{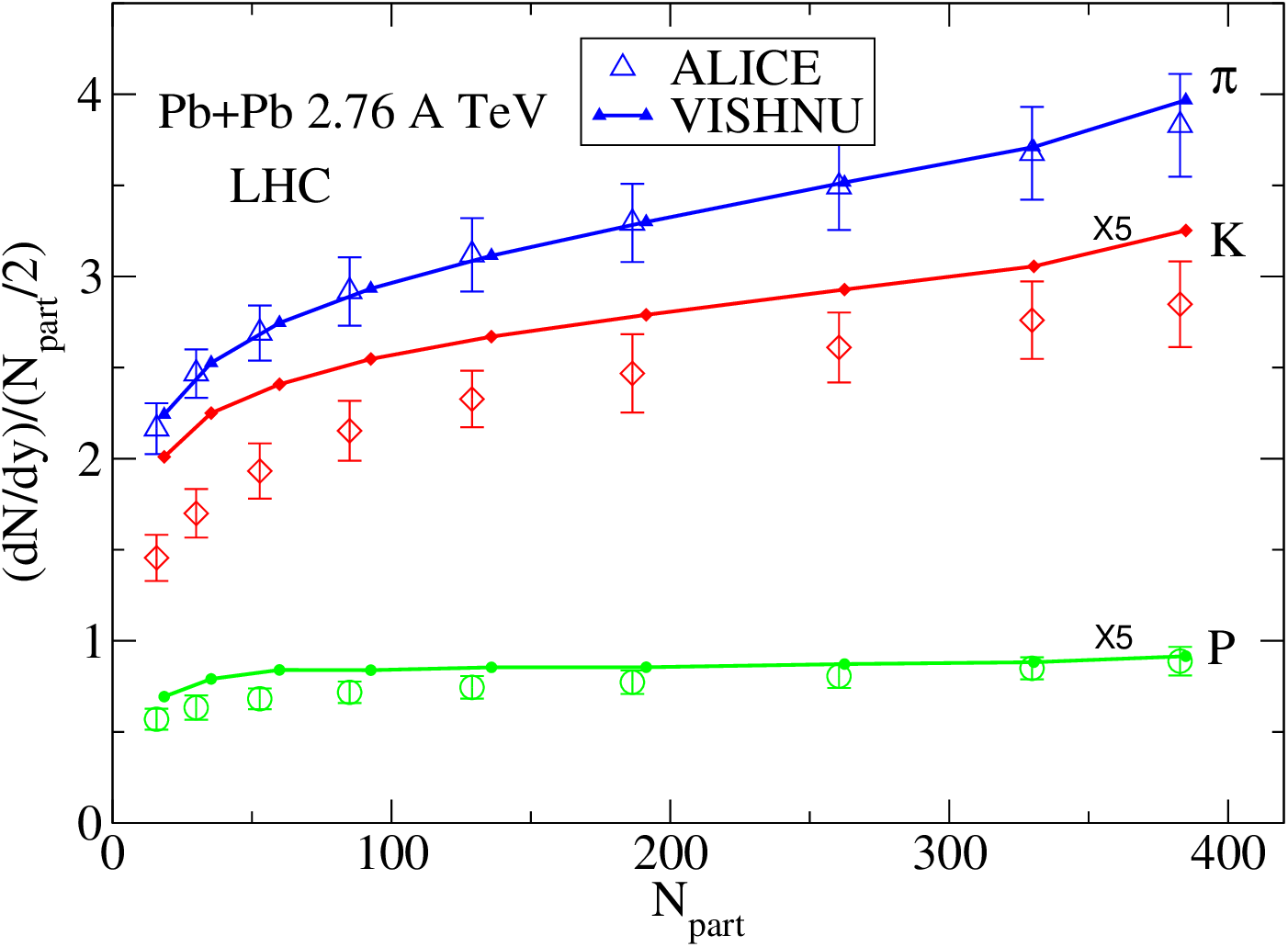}
\caption{(Color online) Centrality dependence of the rapidity density per participant pair $(dN_\mathrm{ch}/dy)/(N_\mathrm{part}/2)$ for pions kaons and protons in 2.76 A TeV Pb+Pb collisions. Experimental data are from ALICE~\cite{Abelev:2013vea}. Theoretical curves are from {\tt VISHNU} calculations with MC-KLN initializations, $\eta/s{\,=\,}0.16$ and
a switching temperature $T_\mathrm{sw}{\,=\,}165$\,MeV.
\label{F1}
}
\end{figure}
%%%%%%%%%%%%%%%%%%%%%%%%%%%%%%%%%%%%%%%%%%%%%%%%%%

%%%%%%%%%%%%%%%% Fig. 2 %%%%%%%%%%%%%%%%%%%%%%%%%%%%%%
\begin{figure*}[tph!]
\includegraphics[width=0.85\linewidth,clip=]{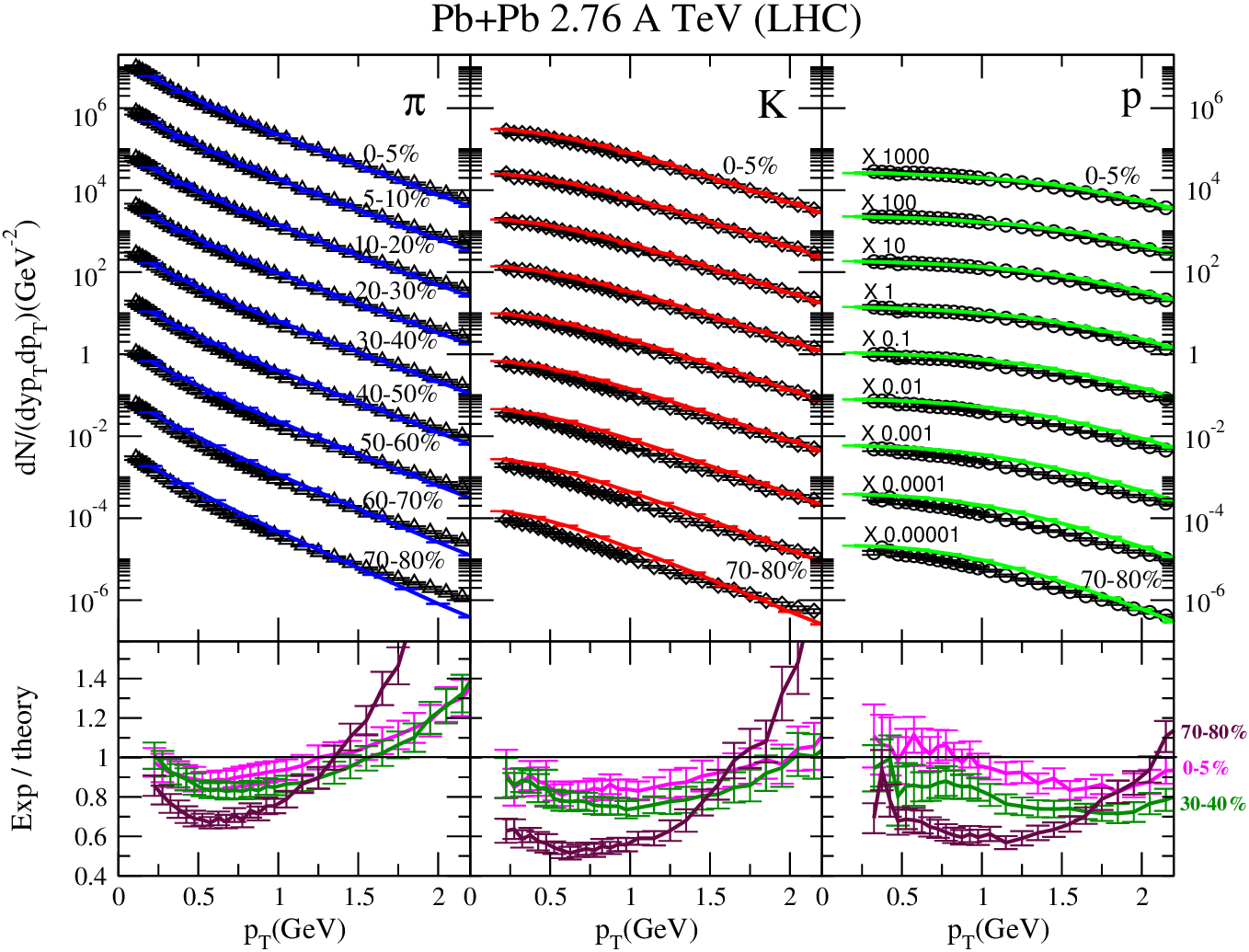}
\caption{(Color online) Upper panels: $p_T$-spectra for pions, kaons and protons in 2.76 A TeV Pb+Pb collisions. Experimental data are from ALICE~\cite{Abelev:2013vea}. Theoretical curves are from {\tt VISHNU}. From top to bottom the curves correspond to 0-5\% ($\times1000$), 5-10\% ($\times100$), 10-20\% ($\times10$), 20-30\%, 30-40\% ($\times0.1$), 40-50\% ($\times0.01$), 50-60\% ($\times0.001$), 60-70\% ($\times10^{-4}$), 70-80\% ($\times10^{-5}$) centrality, respectively, where the factors in parentheses indicate the multipliers applied to the spectra for clearer presentation. Lower panels: the ratio of the experimental and theoretical $p_T$-spectra for pions, kaons and protons for 0-5\%, 30-40\% and 70-80\% centralities.
\label{F2}  }
\end{figure*}
%%%%%%%%%%%%%%%%%%%%%%%%%%%%%%%%%%%%%%%%%%%%%%%%%%

%%%%%%%%%%%%%%%% Fig. 3 %%%%%%%%%%%%%%%%%%%%%%%%%%%%%%
\begin{figure*}[tph!]
\includegraphics[width=0.85\linewidth,clip=]{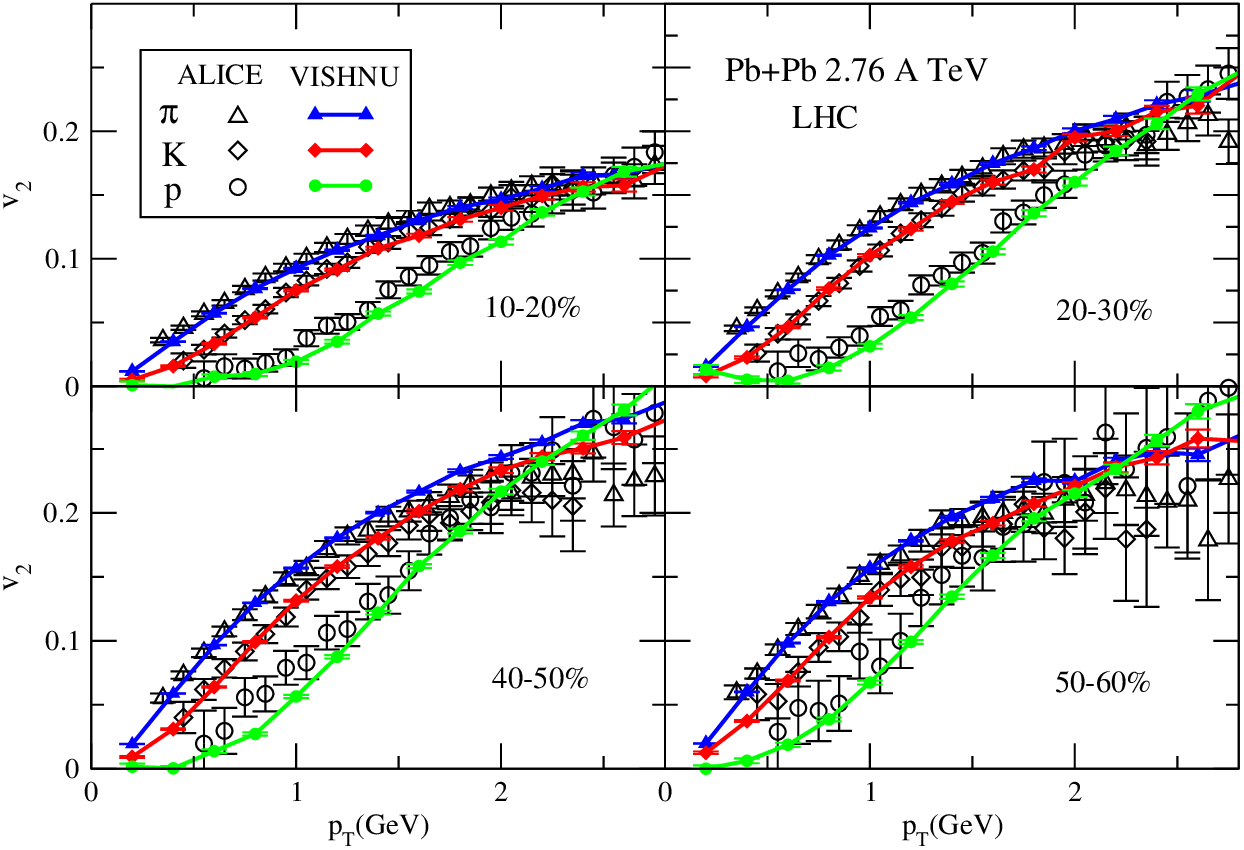}
\caption{(Color online) Differential elliptic flow $v_2(p_T)$ for pions, kaons and protons in 2.76\,$A$\,TeV Pb+Pb collisions. Experimental data are from ALICE~\cite{Noferini:2012ps}. Theoretical curves are from {\tt VISHNU}.  See text for details.
\label{F3}
}
\end{figure*}
%%%%%%%%%%%%%%%%%%%%%%%%%%%%%%%%%%%%%%%%%%%%%%%%%%

In this section we compare our {\tt VISHNU} calculations to the identified hadron multiplicities,
spectra and elliptic flow measurements for pions, kaons and protons that have recently become available for 2.76\,$A$\,TeV Pb+Pb collisions~\cite{Abelev:2013vea}. In our calculations, we use $(\eta/s)_\mathrm{QGP}{\,=\,}0.16$ to describe the elliptic flow data. For a given $(\eta/s)_\mathrm{QGP}$, the initial time $\tau_0$ and the normalization factor for the entropy density are fitted using the charged hadron multiplicity density and $p_T$ spectra for all charged hadrons in the most central collisions. It is found that the curves showing the multiplicity density per participant pair $(dN_\mathrm{ch}/dy)/(N_\mathrm{part}/2)$ vs. participant number $N_\mathrm{part}$ for all charged and identified hadrons are insensitive to the QGP specific shear viscosity with properly tuned $\tau_0$ and normalization parameters~\cite{Song:2010mg,Song:2011hk,HSong2013}. This shows that the centrality dependence of viscous entropy production during the hydrodynamic evolution is weak.

Figure~\ref{F1} shows the centrality dependence of the multiplicity density for identified pions, kaons and protons in 2.76\,$A$\,TeV Pb+Pb collisions. The {\tt VISHNU} hybrid model provides a good description of $(dN^\pi_\mathrm{ch}/dy)/(N_\mathrm{part}/2)$ vs. $N_\mathrm{part}$ for pions over the entire centrality range. It is also capable of describing the proton data in central and semi-central collisions but slightly overpredicts the data in peripheral collisions. Baryon-antibaryon ($B{-}\bar{B}$) annihilation processes in the hadronic phase reduce the proton multiplicity by ${\cal O}(30\%)$; without accounting for $B{-}\bar{B}$ annihilation, the measured proton multiplicities can not be reproduced. This will be studied in more detail in Sec.~\ref{sec5}.  For the LHC data shown in Fig.~\ref{F1}, {\tt VISHNU} overpredicts the kaon multiplicities at all centralities, by about 10\% in central and about 25\% in peripheral collisions. A similar overprediction is also found at RHIC energies (Sec.~\ref{sec5}). $B{-}\bar{B}$ annihilation in {\tt UrQMD} influences kaon production by only ${\cal O}(5\%)$, and hence the discrepancies with the measured kaon yields persist independent of whether or not $B{-}\bar{B}$ annihilation is included. This issue deserves additional investigation.

Figure~\ref{F2} shows the $p_T$ spectra for identified hadrons in 2.76\,$A$\,TeV Pb+Pb collisions. The theoretical lines are calculated with {\tt VISHNU} using the same input parameters as in Fig.~\ref{F1}. The lower panels plot the ratio of experimental and theoretical $p_T$-spectra for the three selected centralities at 0-5\%, 30-40\% and 70-80\%. Except for the most peripheral collisions, where we would not expect the model to perform well, {\tt VISHNU} provides a good description of the ALICE data for all three particle species over most of the centrality range. The calculated $p_T$ spectra for kaons and protons are slightly above the experimental ones, with deviations gradually increasing from central to peripheral collisions, as expected from Fig.~\ref{F1}. In spite of this normalization issue, {\tt VISHNU} is generally capable of correctly describing the slopes of the $p_T$ spectra for these identified hadron species, which reflect the radial flow accumulated in both the QGP and the hadronic phase, over most of the centrality range.

In Figure~\ref{F3} we compare the calculated differential elliptic flow for pions, kaons and proton with experimental data from the ALICE collaboration. The data show $v_2\{\mathrm{SP}\}(p_T)$ which was extracted using the scalar product method~\cite{Noferini:2012ps}. With $(\eta/s)_\mathrm{QGP}{\,=\,}0.16$,\footnote{In the proceedings~\cite{Heinz:2011kt} we used a value of
     $(\eta/s)_\mathrm{QGP}{\,=\,}0.20$, yielding $v_2(p_T)$ values for pions, kaons and
     protons that were slightly lower than the preliminary elliptic flow data reported by ALICE
     at the {\sl Quark Matter 2011} conference \cite{Collaboration:2011yba}. After reducing
     $(\eta/s)_\mathrm{QGP}$ from 0.20 to 0.16, the calculated $v_2(p_T)$ for these
     identified hadrons increased by ${\cal O}(5\%)$, providing an improved description of
     the experimental data.}
{\tt VISHNU} nicely describes the identified hadron elliptic flow data up to 2\,GeV for all shown centralities. The value of $(\eta/s)_\mathrm{QGP}$ used here is slightly below the value $0.20{-}0.24$ used in our earlier work~\cite{Song:2011qa}, which was obtained from fitting the integrated and differential elliptic flow $v_2\{4\}$ for all charged hadrons. The scalar product method flow measurements $v_2\{\mathrm{SP}\}$ use two particle correlations, which are known to over-estimate the mean flow signal due to non-flow contributions and fluctuations. In contrast, the four particle cumulant method $v_2\{4\}$ minimizes non-flow contributions and receives a negative contribution from flow fluctuations, leading to somewhat lower flow values. Due to its larger flow signal compared to $v_2\{4\}$, $v_2\{\mathrm{SP}\}$ therefore leads to a slightly lower value of the extracted QGP shear viscosity $(\eta/s)_\mathrm{QGP}$.\footnote{The reader may correctly object that one should not compare different flow measures in the experimental data and theoretical calculations. Unfortunately, it is difficult to eliminate the effect of flow fluctuations from experimental flow measurements, and including them on the theoretical side requires an event-by-event evolution approach which is prohibitively expensive with the {\tt VISHNU} hybrid code. We therefore emphasize that the analysis presented here does not aim at a precision extraction of $(\eta/s)_\mathrm{QGP}$ -- this would indeed require an event-by-event approach. The goal here is rather to show that we can get a consistent overall description of all soft-hadron observables with a common set of parameters, and use this to make predictions for so far unpublished measurements of additional hadron species.}

The authors of \cite{Shen:2011eg} previously predicted the elliptic flow for pions, kaons and protons at the LHC using a pure (2+1)-d viscous hydrodynamic calculation that employed the fluid dynamic code {\tt VISH2{+}1} to describe the evolution of both the QGP and hadronic phases. With their choice of parameters, an MC-KLN initialization, a constant value of $\eta/s{\,=\,}0.20$, and a decoupling temperature $T_\mathrm{dec}{\,=\,}120$\,MeV, they nicely predicted the later shown ALICE data \cite{Collaboration:2011yba} for $v_2(p_T)$ below $p_T{\,<\,}1.5$\,GeV for pions and kaons for mid-central to mid-peripheral centralities bins. However, since the calculation assumed chemical freeze-out at $T_\mathrm{chem}{\,=\,}165$\,MeV and ignored $B{-}\bar{B}$ annihilation below $T_\mathrm{chem}$, they over-predicted the proton yields. The {\em shapes} of the proton $p_T$ spectra were predicted reasonably well over most of the measured centrality range, except for the most central collisions where the predicted spectra lacked radial flow and were somewhat too steep. This latter problem also affected the differential flow of protons, $v_2^p(p_T)$, which {\tt VISH2{+}1} overpredicted in central to semi-central collisions below 2\,GeV. The microscopic hadronic rescattering processes in {\tt UrQMD}, contained in the {\tt VISHNU} hybrid model employed here, rebalance the generation of radial and elliptic flow for protons, leading to an improved description of the proton $p_T$ spectra and $v_2(p_T)$ in central and semi-central collisions; the inclusion in {\tt VISHNU} of $B{-}\bar{B}$ annihilation below $T_\mathrm{chem}$ corrects the problem with the proton yields from the pure {\tt VISH2{+}1} approach. The elliptic flow for pions and kaons are equally well described  in both {\tt VISH2+1} and {\tt VISHNU}. Compared to $(\eta/s)_\mathrm{QGP}=0.16$ used in the {\tt VISHNU} calculation, {\tt VISH2+1} used a somewhat larger $\eta/s$ value of 0.2 for the combined QGP and hadronic evolution. Apparently this successfully compensates for the larger dissipative effects in the hadronic phase that are insufficiently described in a purely hydrodynamic approach. A detailed {\tt UrQMD} analysis shows that protons decouple from the system later than pions and kaons~\cite{Song-SQM2013}. This explains why a uniform freeze-out temperature of 120 MeV (used for pions and kaons) fails to describe the proton data in the pure hydrodynamic approach (at least in central to mid-central collisions).

%============================ Fig. 4 ==================================
\begin{figure}[tph]
\includegraphics[width=0.85\linewidth,clip=]{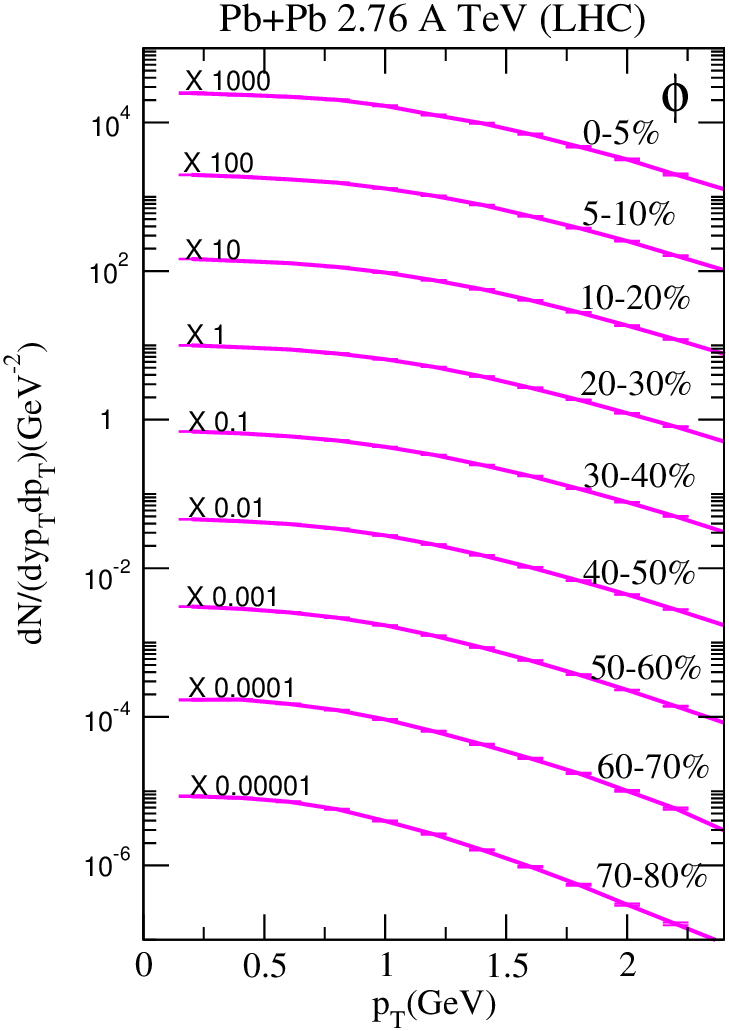}
\caption{(Color online) A prediction for the $\phi$ meson $p_T$ spectra for 2.76\,$A$\,TeV Pb+Pb collisions at different centralities as indicated, using {\tt VISHNU} simulations with the same inputs as in Figs.~\ref{F1}-\ref{F3}.
\label{F4}
}
\end{figure}
%======================================================================

%%%%%%%%%%%%%%%%%%%%%%%%%%%%%%%%%%%%%%%%%%%%%%%%%
\section{$\bm{p_T}$ spectra and elliptic flow of $\bm{\phi}$ mesons}\label{sec4}

%============================ Fig. 5 ==================================
\begin{figure*}[tph]
\includegraphics[width=0.85\linewidth,clip=]{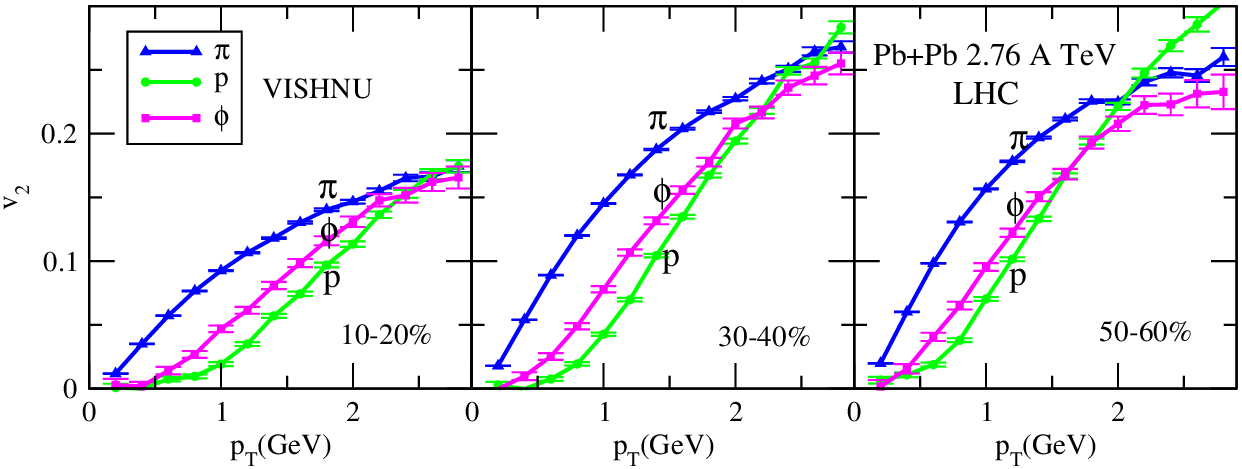}
\caption{(Color online) A prediction for the $\phi$ elliptic flow for 2.76\,$A$\,TeV Pb+Pb collisions  using {\tt VISHNU} simulations with the same inputs as in Figs.~\ref{F1}-\ref{F3}.
\label{F5}
}
\end{figure*}
%======================================================================
%============================ Fig. 6 ==================================
\begin{figure*}[tph]
\includegraphics[height=6.13cm,width=0.4\linewidth,clip=]{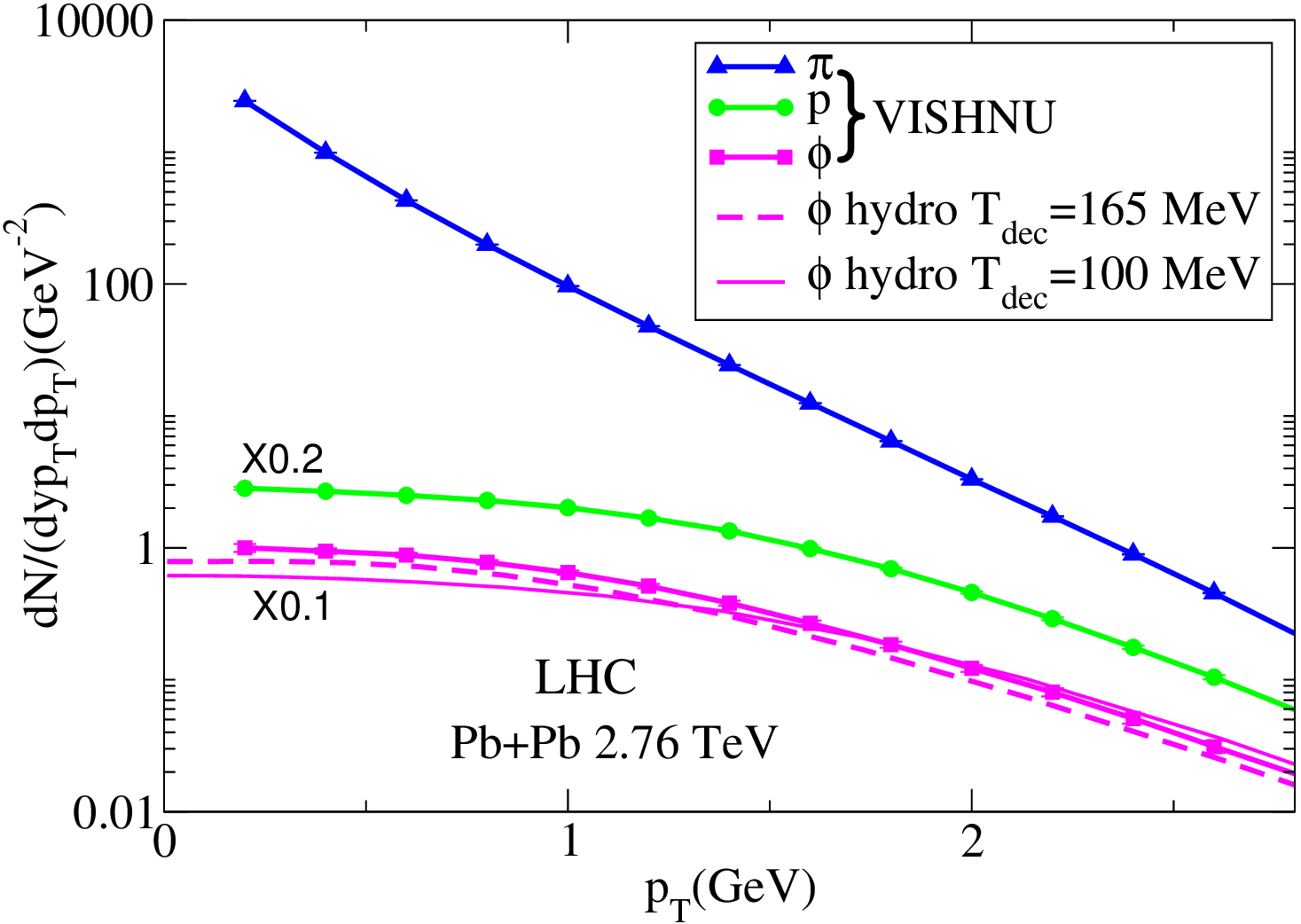}
\includegraphics[height=6cm,width=0.4\linewidth,clip=]{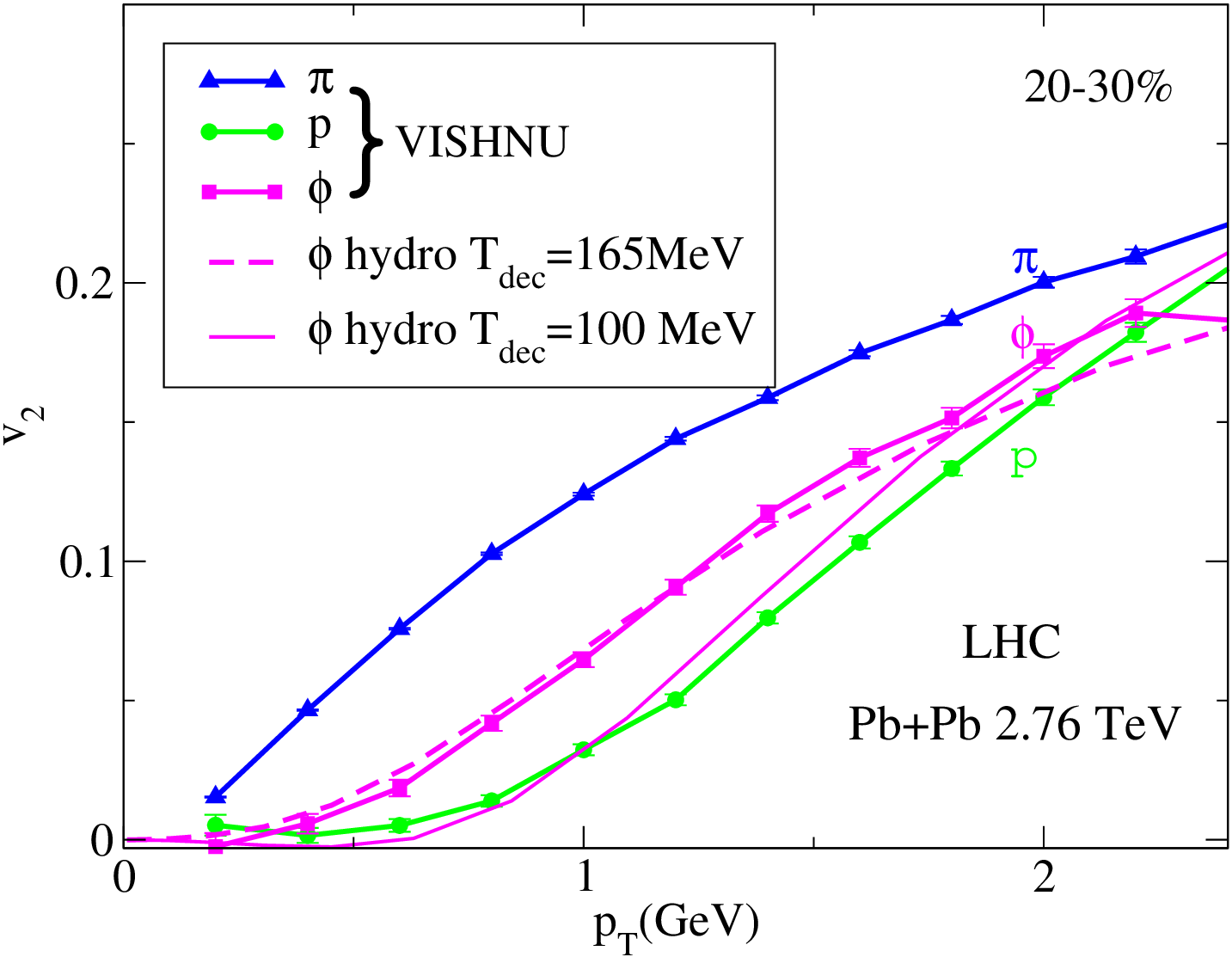}
\caption{(Color online) $p_T$ spectra and differential elliptic flow $v_2(p_T)$ for pions, protons and $\phi$ mesons from the hybrid code {\tt VISHNU}, compared with results for $\phi$ mesons from pure viscous hydrodynamics {\tt VISH2+1} with resonance decays. The freeze-out temperature for {\tt VISH2+1} is set to 165 MeV and 100 MeV, respectively, and the switching temperature for {\tt VISHNU} is 165 MeV. See text for details.
\label{F6}
}
\end{figure*}
%======================================================================

Building upon our successful description of the $p_T$ spectra and elliptic flow of pions, kaons and protons, we now focus on the elliptic flow of the $\phi$ meson at LHC energies using the {\tt VISHNU} hybrid approach. Fig.~\ref{F4} shows a prediction for the $\phi$ meson $p_T$ spectra corresponding to the pion, kaon, and proton $p_T$ spectra shown in Fig.~\ref{F2}. Even though in {\tt UrQMD} $\phi$ mesons are affected significantly less by hadronic rescattering than protons (see discussion below), their $p_T$ spectra exhibit a clear ``flow shoulder", similar to that seen for protons in Fig.~\ref{F2}. This shows that a large fraction of the finally observed radial flow is already created in the QGP phase.

In Fig.~\ref{F5} we compare the differential elliptic flow $\phi$ mesons in 2.76 A TeV Pb+Pb collisions to that of pions and protons (shown already in Fig.~\ref{F3}). In the {\tt VISHNU} calculation $v_2^{\phi}(p_T)$ runs above the proton $v_2^{p}(p_T)$ curve for $p_T{\,<\,}1.5{-}2$\,GeV, but drops below at higher $p_T$. This result agrees qualitatively with predictions by Hirano {\it et al.} on the mass-ordering between proton and $\phi$ elliptic flow in 200\,$A$\,GeV Au+Au collisions, using a hybrid model that couples (3+1)-dimensional ideal hydrodynamics with the JAM hadron cascade \cite{Hirano:2007ei}. In spite of the different collision energies and centralities and other differences in the two hybrid model simulations (viscosity, longitudinal dynamics, EOS and initializations), both calculations agree in the prediction that the $\phi$ meson elliptic flow violates the traditional hydrodynamic mass-ordering, due to a smaller rescattering cross-section in the hadron gas evolution than for protons which results in an earlier decoupling of the $\phi$ from the buildup of additional radial flow in the hadronic phase \cite{Hirano:2007ei}.

Preliminary data reported by the STAR Collaboration \cite{Nasim:2012gz} for $\phi$ meson elliptic flow $v_2^{\phi}(p_T)$ in 200\,$A$\,GeV Au+Au collisions at RHIC confirmed the predicted mass-ordering violation in the region $p_T{\,<\,}1$\,GeV while recovering standard mass-ordering at higher $p_T$. However, the crossing between the curves for $v_2^{\phi}(p_T)$ and $v_2^{p}(p_T)$ observed in \cite{Nasim:2012gz} happens at a lower $p_T$ value, and the splitting between proton and $\phi$ elliptic flow at higher $p_T$ is found to be significantly larger than predicted by Hirano {\it et al.} for RHIC energies in \cite{Hirano:2007ei}, and confirmed here in Fig.~\ref{F5} for LHC energies. Forthcoming results from a measurement of $v_2^{\phi}(p_T)$ in 2.76\,$A$\,TeV Pb+Pb collisions at the LHC are expected shed further light on this matter.

To explore in greater depth the dynamical evolution of the $\phi$ meson spectra and elliptic flow in the hadronic stage we compare in Fig.~\ref{F6} $v_2^\phi$ calculated by {\tt VISHNU} to that obtained in pure viscous hydrodynamics with decoupling temperatures set to 165 MeV and 100 MeV, respectively. $T{\,=\,}165$\,MeV is the switching temperature in {\tt VISHNU} for the transition from hydrodynamic to microscopic evolution; setting $T_\mathrm{dec}{\,=\,}T_\mathrm{sw}$ thus completely eliminates the hadronic stage from the flow evolution and therefore corresponds to the limiting case of an infinitely weakly coupled hadron gas stage. The choice $T_\mathrm{dec}{\,=\,}100$\,MeV, on the other hand, assumes the validity of hydrodynamics all the way to a very low kinetic freeze-out density and thus implements the opposite extreme of a very strongly coupled hadron gas phase.\footnote{In principle, we could play with the shear viscosity in
     the hadronic phase to explore different coupling strengths, but we found that the
     code {\tt VISH2+1} develops numerical instabilities when changing $\eta/s$
     discontinuously at the switching temperature. We therefore continued the hadronic
     evolution with the same specific shear viscosity $\eta/s{\,=\,}0.16$ used in the QGP
     phase, corresponding to very, but not infinitely  strong coupling in the hadronic phase.}
Figure~\ref{F6} shows that the {\tt VISHNU} calculation including the microscopic hadron gas evolution with finite cross-sections for all hadron species, which should provide the most  realistic description of the hadron gas dynamics, yields $\phi$ meson $p_T$ spectra with slopes close to the hydrodynamic curves for $T_\mathrm{dec}{\,=\,}165$\,MeV. The ${\cal O}(20\%)$ larger $\phi$ yield in the {\tt VISHNU} calculation compared to {\tt VISH2{+}1} arises from additional $\phi$ production via $K^+K^-$ scattering in {\tt UrQMD} while $\phi$ meson decays are turned off (otherwise no $\phi$ mesons would be left at the end of the {\tt UrQMD} stage). The $\phi$ yield shown in Fig.~\ref{F6} should be experimentally accessible via the dilepton decay channel of the $\phi$, whereas a measurement through its hadronic decay channel will yield a lower yield since $\phi\to K^+K^-$ decays where one of the kaons rescatters in the hadronic phase will lead to a loss of reconstructed $\phi$ mesons.

The elliptic flow of phi mesons lies between the strongly and weakly coupled hadron gas limits, but closer to the latter. This confirms that, due to its small hadronic cross sections as implemented in {\tt UrQMD}, the $\phi$ meson is rather weakly coupled to the hadronic medium and tends to decouple from the system almost directly after hadronization, without significant further interactions. Note that neither the hybrid code {\tt VISHNU} nor pure hydrodynamics with $T_\mathrm{dec}{\,=\,}100$\,MeV are able to produce a $\phi$ meson elliptic flow that lies significantly below the proton elliptic flow for transverse momenta between 1 and 2\,GeV. It will be interesting to see whether upcoming experimental analyses support this prediction.

%%%%%%%%%%%%%%%%%%%%%%%%%%%%%%%%%%%%%%%%%%%%%%%%%
\section{$B$-$\bar{B}$ annihilation and soft hadron production in VISHNU}\label{sec5}

%============================ Fig. 7 ==================================
\begin{figure*}[tph]
\includegraphics[width=0.45\linewidth,clip=]{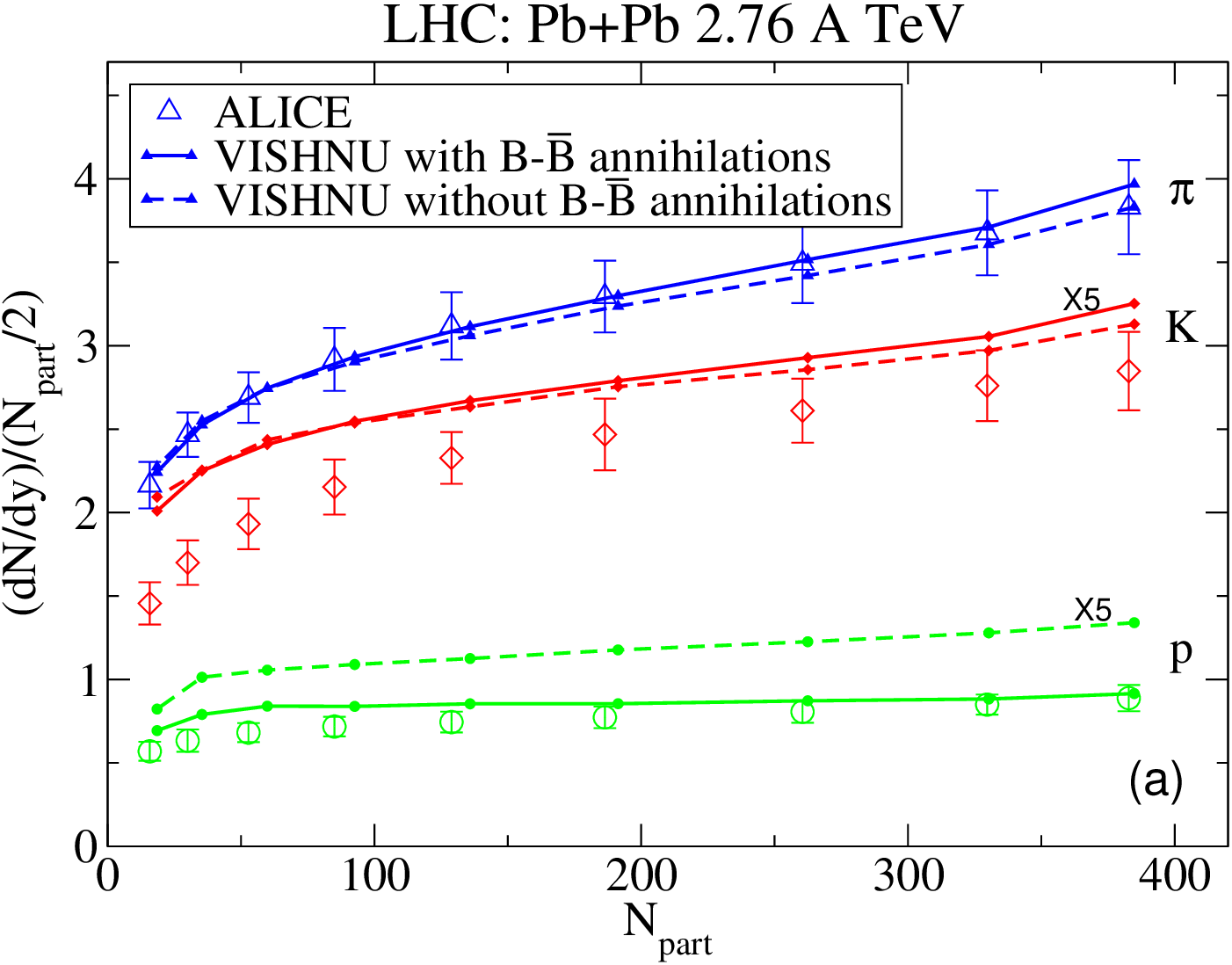}
%\vspace{30mm}
\includegraphics[width=0.475\linewidth,clip=]{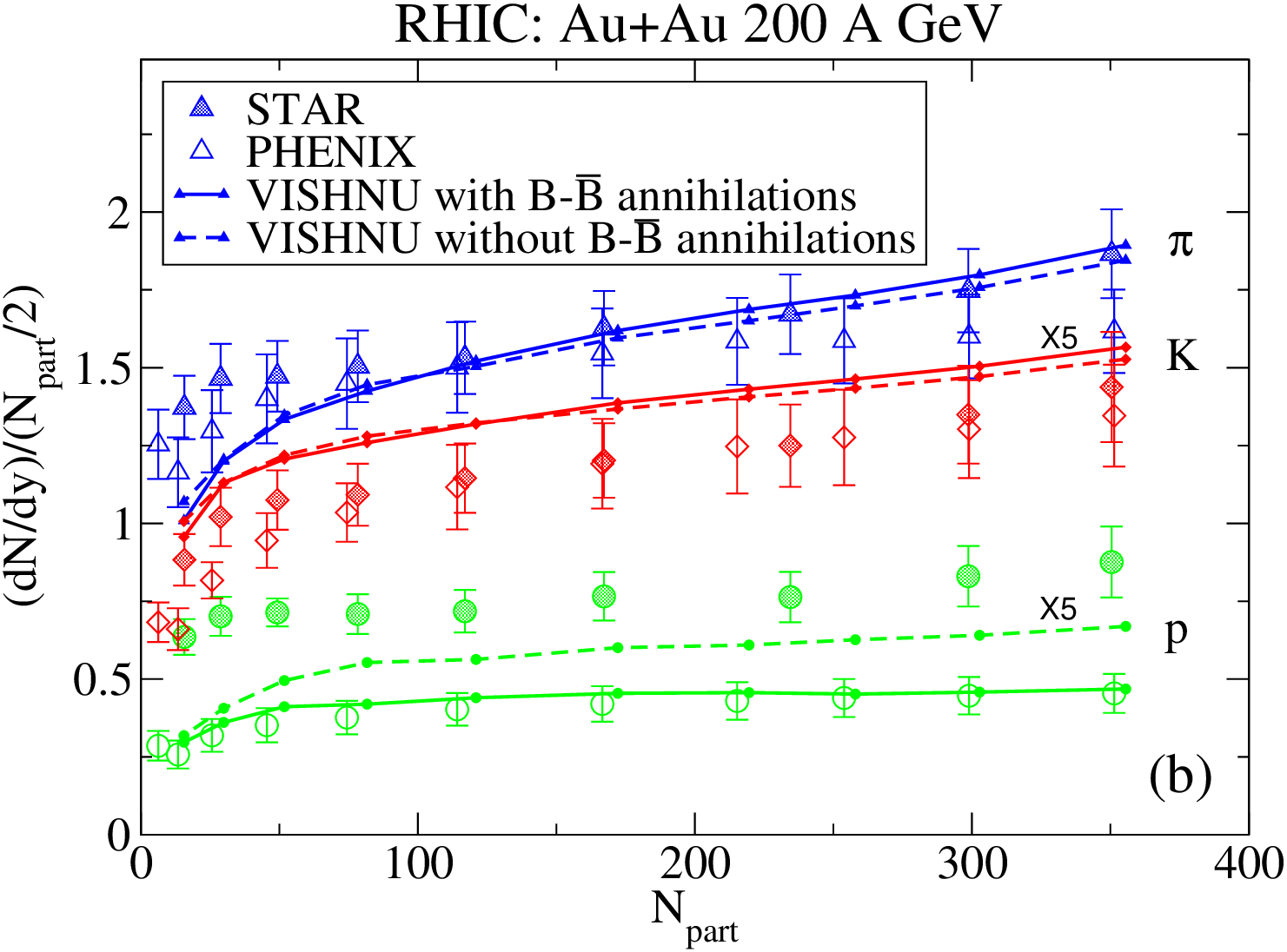}
\caption{(Color online) Centrality dependence of the rapidity density per participant pair $(dN_\mathrm{ch}/dy)/(N_\mathrm{part}/2)$ for pions kaons and protons in 2.76\,$A$\,TeV Pb+Pb collisions (left) and in 200\,$A$\,GeV Au+Au collisions (right). Experimental data are from ALICE~\cite{Abelev:2013vea}, STAR~\cite{:2008ez}, and PHENIX~\cite{Adler:2003cb}. Theoretical curves are from VISHNU, with $B{-}\bar{B}$ annihilation turned on (solid lines) or off (dashed lines). The left panel is similar to Fig.~\ref{F1}, except for the addition of the theoretical lines without $B{-}\bar{B}$ annihilation.
\label{F7}
}
%\end{figure*}
\end{figure*}
%======================================================================

In the early version of {\tt VISHNU} used in~\cite{Song:2010mg,Song:2011hk,Song:2011qa}, the baryon-antibaryon annihilation processes in the hadronic Boltzmann transport {\tt UrQMD} were accidentally turned off. It was found that these $B$-$\bar{B}$ annihilation channels could significantly reduce the proton and anti-proton multiplicities by ${\cal O}(30\%)$. In the errata of Refs.~\cite{Song:2010mg,Song:2011hk,Song:2011qa}, we re-calculated the corresponding spectra and $v_2$ figures including $B{-}\bar{B}$ annihilation processes, but did not directly compare our results with and without $B{-}\bar{B}$ annihilation processes. We here focus on the study of soft physics for identified hadrons, where the multiplicities and spectra for pions, kaons and protons may all be affected to a varying degree by $B{-}\bar{B}$ annihilations during the hadronic evolution.

\begin{figure*}[tph]
%\vspace{0.2in}
\includegraphics[width=0.45\linewidth,clip=]{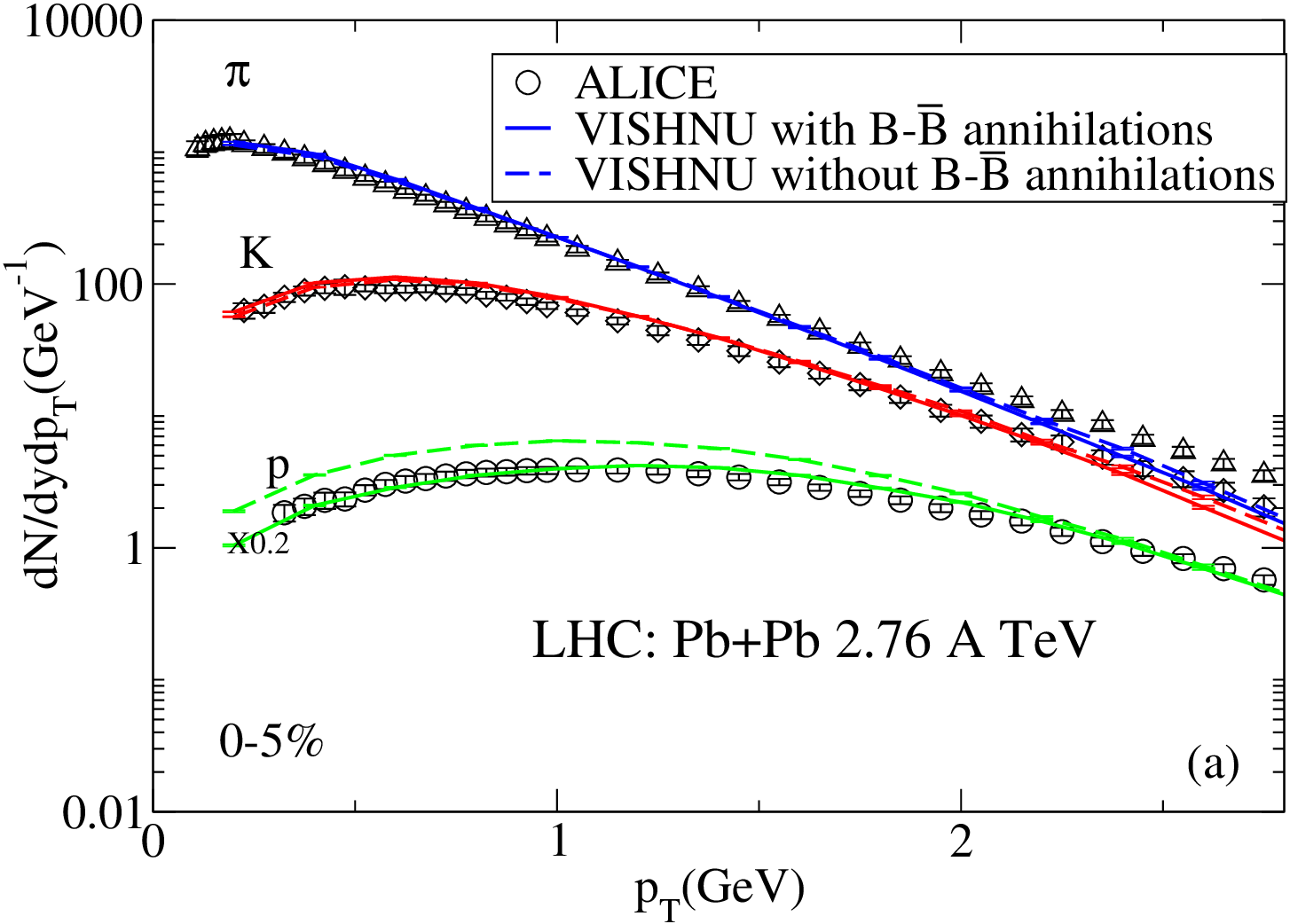}
\includegraphics[width=0.45\linewidth,clip=]{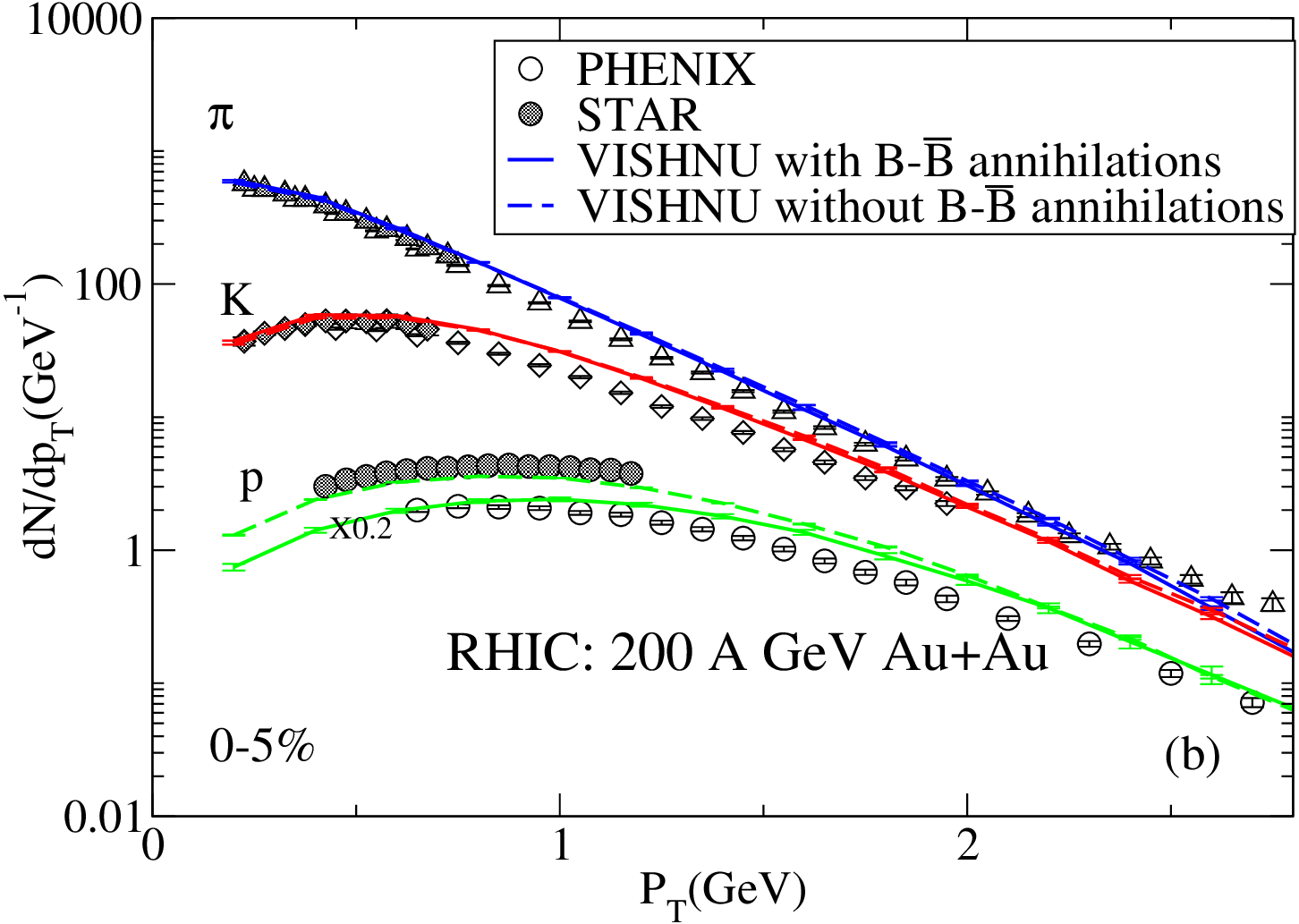}
\caption{(Color online) $p_T$ spectra in most central collisions for pions kaons and protons in 2.76\,$A$\,TeV Pb+Pb collisions (left) and 200\,$A$\,GeV Au+Au collisions (right). Experimental data are from from ALICE~\cite{Abelev:2013vea}, STAR~\cite{:2008ez,Adams:2003kv}, and PHENIX~\cite{Adler:2003cb}, respectively. Same illustration for theoretical and experimental lines as in Fig.~\ref{F7}. All proton spectra were divided by 5 for improved clarity of the plots.
\label{F8}
}
\end{figure*}

Figure~\ref{F7} shows the multiplicity density per participant pair $(dN_\mathrm{ch}/dy)/(N_\mathrm{part}/2)$ for pions, kaons and protons at RHIC and LHC. The lines denote {\tt VISHNU} calculations with or without activation of $B{-}\bar{B}$ annihilation channels. $B{-}\bar{B}$ annihilation processes mainly affect the multiplicities of protons and anti-protons; they reduce $dN_p/dy$ and $dN_{\bar{p}}/dy$ by ${\cal O}(30\%)$.\footnote{At zero net baryon density,
    due to the isospin symmetry in {\tt UrQMD}, the final multiplicity, spectra and flow for
    protons and anti-protons are identical if the number of events is chosen sufficiently large.} When including annihilation processes, {\tt VISHNU} provides a good description of the proton multiplicities over the entire centrality range, as measured by PHENIX~\cite{Adler:2003cb} at RHIC and by ALICE~\cite{Abelev:2013vea} at the LHC.\footnote{Note that the STAR data
    \cite{:2008ez} in Figs.~\ref{F7} and \ref{F8} include protons from hyperon decays with
    a detection efficiency that we cannot easily simulate in {\tt VISHNU}. These extra protons
    from weak decays account for the ${\cal O}(50\%)$ difference between the STAR data
    and those from the PHENIX experiment \cite{Adler:2003cb} from which protons from
    weak decays have been removed. Our {\tt VISHNU} results do not include any weak decay
    protons.}
In addition to the normalization, inclusion of $B{-}\bar{B}$ annihilation also improves the {\em shape} of the proton $p_T$ spectra at the LHC when compared with the ALICE data (shown in Fig.~\ref{F8}a). (Note that the ALICE data have been corrected for feed-down from weak decays \cite{Abelev:2013vea}.) At RHIC energies (Fig.~\ref{F8}b), $B{-}\bar{B}$ annihilation improves the normalization of the spectra in comparison with the PHENIX data, but renders the slope of the proton spectra somewhat flatter than seen in the experiment.

$B{-}\bar{B}$ annihilation processes mostly impact low $p_T$ baryon and antibaryon multiplicities, leading to a suppression and a softening of the proton $p_T$ spectra below 2 GeV. However, they also produce additional mesons, which causes a $\sim 4\%$ increase in the pion multiplicity and a $\sim 2\%$ increase in the kaon multiplicity.  To compensate for this added particle multiplicity, we had to reduce the normalization factor of the entropy density in our initial condition by $\sim 4\%$, in order to retain the previous good overall description of the charged hadron multiplicities. The solid lines in Figs.~\ref{F7} and \ref{F8} correspond to the {\tt VISHNU} results with modified initial conditions, showing a significant reduction in the number of protons and a slight increase in the number of pions and kaons. The effects of $B{-}\bar{B}$ annihilation are more prominent in the most central collisions and at higher collision energies, due to the longer evolution time in the hadronic stage. Correspondingly, the suppression of the proton multiplicity is smaller at RHIC than at the LHC, and slightly decreases from central to peripheral collisions as shown in Fig.~\ref{F7}. The resulting decrease in the initial entropy density, if not corrected as we have done here, leads to a slightly reduced value for the integrated $v_2$, as noted in the errata Refs.~\cite{Song:2011hk,Song:2011qa}. Within current statistics, the differential $v_2$ for identified hadrons are not noticeably affected by the $B{-}\bar{B}$ annihilation channels.

%============================ Fig. 8 ==================================
%======================================================================

The  multiplicities for various hadron species at RHIC and the LHC have also been studied within the framework of the Statistical Model.  Using a chemical freeze-out temperature $T_\mathrm{ch}{\,=\,}164$\,MeV extracted from hadron yields measured by STAR in 200\,$A$\,GeV Au+Au collisions \cite{Adams:2005dq}, the Statistical Model over-predicts the proton and anti-proton yields observed in 2.76\,$A$\,TeV Pb + Pb collisions \cite{Andronic:2012dm}. In that framework, in order to obtain a better description of the proton and anti-proton data, the chemical freeze-out temperature at the LHC  needs to be reduced to $\sim$155\,MeV. However, recent hybrid hydro+micro model calculations have demonstrated that the out-of equilibrium evolution of the system during the hadronic phase plays an important role for a proper description of the proton and anti-proton data~\cite{Steinheimer:2012rd,Song:2012ua}. Our calculations confirm these findings. In this section, we have shown that {\tt VISHNU} calculations that include $B{-}\bar{B}$ annihilation can simultaneously describe the soft pion, kaon and proton production measured by the PHENIX and ALICE collaborations at top RHIC and the LHC energies, using a switching temperature of 165 MeV at which {\tt UrQMD} is initialized with chemical equilibrium abundances. The demonstrated suppression of final $B$ and $\bar{B}$ yields by inelastic collisions, including annihilation processes, during the {\tt UrQMD} rescattering stage demonstrates that proton and antiproton yields effectively freeze out below $T_\mathrm{sw}$, and that the final chemical composition therfore cannot be accurately described by a single chemical freeze-out temperature. In a future study using larger event statistics we plan to explore the relative importance of similar $B{-}\bar{B}$ annihilation processes on the strange and multi-strange hyperon yields.

%%%%%%%%%%%%%%%%%%%%%%%%%%%%%%%%%%%%%%%%%%%%%%%%%
\section{Summary and Outlook}\label{sec6}

In this article, we have studied  soft particle production in 2.76\,$A$\,TeV Pb+Pb collisions using the {\tt VISHNU} hybrid model that describes the expansion of the viscous QGP fluid with a hydrodynamic model and the successive evolution of the hadronic gas with a microscopic hadron transport model. Using {\tt MC-KLN} initial conditions, a value of $(\eta/s)_\mathrm{QGP}=0.16$ and a switching temperature $T_\mathrm{sw}=165$\,MeV, {\tt VISHNU} provides a good description of identified hadron multiplicities, $p_T$ spectra and differential elliptic flow for pions, kaons and protons at various centralities. We used the same calculations to predict the $p_T$-spectra and elliptic flow of $\phi$ mesons. We explored the mass-ordering between the $\phi$ meson and proton differential elliptic flows by comparing the {\tt VISHNU} calculation to the weak and strong coupling limits of the hadron resonance gas phase, simulated with pure hydrodynamics using decoupling temperatures of 165 and 100 MeV, respectively. We investigated the effects of baryon-antibaryon annihilation processes on soft particle production, and showed that, when annihilation processes are included, {\tt VISHNU}  can simultaneously reproduce the multiplicities and $p_T$ spectra for pions, kaons and protons at RHIC and LHC. We discussed the nature of the switching temperature $T_\mathrm{sw}$ in hybrid model calculations and that it cannot be identified as the chemical freeze-out temperature of the system, since inelastic and annihilation processes are still driving the dynamics of the system in its early hadronic evolution. In future work, it may be worthwhile to attempt an extraction of effective chemical freeze-out temperatures for the different hadron species during their evolution in the hadronic phase.

\vspace*{2mm}

\acknowledgments
\vspace*{-2mm}
This work was supported by the new faculty startup funding to H.S. by Peking University and by the U.S.\ Department of Energy under grants \rm{DE-FG02-05ER41367}, \rm{DE-SC0004286}, and (within the framework of the Jet Collaboration) \rm{DE-SC0004104}. We gratefully acknowledge extensive computing resources provided to us by the Ohio Supercomputer Center and on Tianhe-1A by the National Supercomputing Center in Tianjin, China.

%%%%%%%%%%%%%%%%%%%%%%%%%%%%%%%%%%%%%%%%%%%%%%%%%%%%%%%%%%%%%%%%%%%%%%%%%%%%

\end{document}